\documentclass[twocolumn]{cinc}
\usepackage{graphicx, float}

\usepackage{array}
\usepackage{url}
\usepackage{multirow}
\usepackage{silence}
\usepackage{subcaption,booktabs}

\usepackage{float}
\usepackage{hyperref}
\usepackage{mathtools}
\usepackage{xcolor}

\hypersetup{
colorlinks,
citecolor=green
}

\begin{document}

\bibliographystyle{cinc}

\title{A Multi-Task Cross-Task Learning Architecture for Ad-hoc Uncertainty Estimation in 3D Cardiac MRI Image Segmentation}

\author {\textbf{S. M. Kamrul Hasan$^{1}$}\thanks{\rule{4cm}{0.05pt}\\ Corresponding author: \href{https://smkamrulhasan.github.io/}{\color{blue}{S. M. Kamrul Hasan}} (\href{mailto:sh3190@rit.edu}{\color{blue}{Email: sh3190@rit.edu}}). Code, pretrained models, and additional details are available at \href{https://github.com/smkamrulhasan/MTCTL}{\color{red}{https://github.com/smkamrulhasan/MTCTL.}}}, \textbf{Cristian A. Linte$^{1, 2}$} \\
\ \\ % leave an empty line between authors and affiliation
 $^1$ Chester F. Carlson Center for Imaging Science, $^2$  Department of Biomedical Engineering  \\
 Rochester Institute of Technology, Rochester, NY}

\maketitle

\begin{abstract}
Medical image segmentation has significantly benefitted thanks to deep learning architectures. Furthermore, semi-supervised learning (SSL) has recently been a growing trend for improving a model’s overall performance by leveraging abundant unlabeled data. Moreover, learning multiple tasks within the same model further improves model generalizability. To generate smooth and accurate segmentation masks from 3D cardiac MR images, we present a Multi-task Cross-task learning consistency approach to enforce the correlation between the pixel-level (segmentation) and the geometric-level (distance map) tasks. Our extensive experimentation with varied quantities of labeled data in the training sets justifies the effectiveness of our model for the segmentation of the left atrial cavity from Gadolinium-enhanced magnetic resonance (GE-MR) images. With the incorporation of uncertainty estimates to detect failures in the segmentation masks generated by CNNs, our study further showcases the potential of our model to flag low-quality segmentation from a given model.
\end{abstract}

\section{Introduction}
While deep learning has shown its potential in a variety of medical image analysis problems including segmentation \cite{hasan2020co}, motion estimation \cite{zheng2019explainable} etc., many of these successes are achieved at the cost of a large pool of labeled datasets. Obtaining labeled images however is laborious as well as costly, making the adoption of large-scale deep learning models in clinical settings difficult. To address the limited labeled data problem, semi-supervised learning (SSL) \cite{ouali2020semi} has been a growing trend for improving the deep learning model performance through utilizing the unlabeled data. Furthermore, multi-task learning (MTL) \cite{caruana1997multitask} techniques have shown promising results for improving the generalizability of any models by jointly tackling multiple tasks through shared representation learning \cite{zhang2018learning}. To date, a number of approaches address SSL along with MTL-based segmentation from MRI including adversarial learning-based method \cite{hung2018adversarial}, mutual learning-based approach \cite{zhang2021dual} and techniques based on signed distance map \cite{dangi2019distance}. Recent approaches involve integrating uncertainty map into a mean-teacher framework to guide student network \cite{yu2019uncertainty} for left atrium segmentation. However, this method lacks the geometric shape of semantic objects, leading to poor segmentation at the edges. Li {\it et al.} \cite{li2020shape} proposed an adversarial-based decoder to enforce the consistency between the model predictions on the original data and the data perturbed by adding noise into it. 

\begin{figure}[t!]
%\centering
\includegraphics[width=1.0\linewidth]{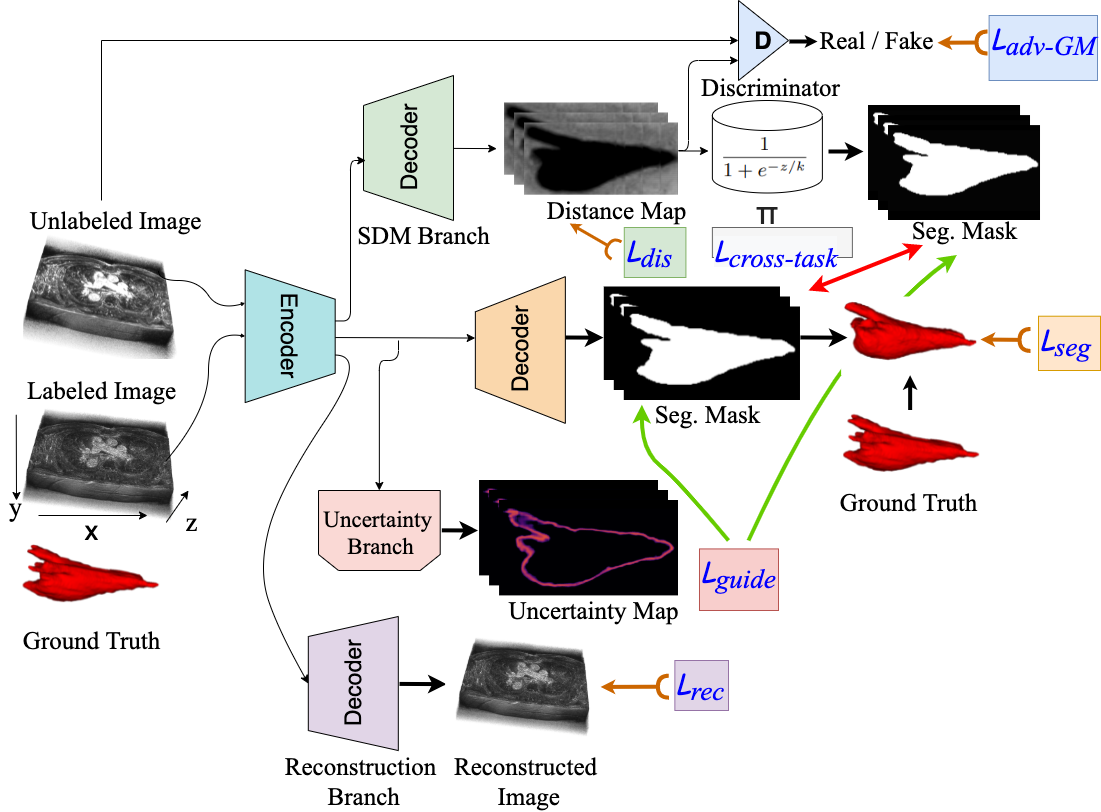}
\caption{Schematic of the \textit{MTCTL} model: we combine four different decoders who share the same backbone encoder -- V-Net.}
\label{FIGURA1}
\end{figure}

% Although, the cardiac MRI has provided a non-invasive method for studying global and regional function of the heart, most of these studies have been centered on only the segmentation and there have only been very few research endeavors exploring the ambiguous predictions in some challenging regions generated by the deep learning models due to the lack of training data.

As a departure from the existing SSL and MTL models, we propose a novel semi-supervised module exploiting adversarial learning and task-based consistency regularization for jointly learning multiple tasks in a single backbone module – uncertainty estimation, geometric shape generation, and cardiac anatomical structure segmentation. The network takes as input a 3D volume and outputs an uncertainty map, a 3D distance map, and a segmentation map. The distance map is fed to a transformer to produce a segmentation map which is then used to share the supervisory signal from the predicted segmentation map. To leverage the unlabeled data, the distance map is fed to an adversarial discriminator network to distinguish the predicted distance map from the labeled data. The same encoder backbone is used to estimate the uncertainty of the predicted segmentation map with Monte Carlo sampling.

\section{Multi-Task Cross-Task Learning}
As shown in Figure \ref{FIGURA1}, our proposed MTCTL model has two distinctive features. First, we combine four different decoders who share the same backbone encoder -- V-Net \cite{milletari2016v}. The uncertainty map generated by the uncertainty decoder is used as the local guidance between the predicted segmentation mask and the mask generated by transforming the distance map. Second, we enforce the correlation between the pixel-level (segmentation) and the geometric-level (distance map) tasks for the generation of smoother and more accurate segmentation masks by introducing the cross-task loss function and include a guidance loss as an uncertainty estimation to smooth out the predicted segmentation mask.

We define the learning task as follows: given an (unknown) data distribution $p(x, y )$ over images and segmentation masks, we have a source domain having a training set, $\mathcal{D_L} =$ $\{(x^l_1, y_1),..., (x^l_n, y_n\}$ with $n$ labeled data and another domain having a training set, $\mathcal{D_{UL}} =$ $\{x^{ul}_1,...,x^{ul}_m\}$ with $m$ unlabeled data which are sampled i.i.d. from $p(x, y)$ and $p(x)$ distribution. Empirically, we want to minimize the target risk $\in_{t}(\phi, \theta) = min_{\phi, \theta} \hspace{3mm} \mathcal{L_L}(\mathcal{D_L}, (\phi, \theta)) + \gamma \mathcal{L_{UL}}(\mathcal{D_{UL}}, (\phi, \theta))$, where $\mathcal{L_L}$ is the supervised loss for segmentation,  $\mathcal{L_{UL}}$ is unsupervised loss defined on unlabeled images and $\phi, \theta$ denotes the learnable parameters of the overall network. 

In this work, our architecture is composed of a shared encoder $e$ and a main decoder $d$, which constitute the segmentation network $f = d \circ e$. We introduce a set of $J$ auxiliary decoders $d_a^j$, with $j \in [1, J]$.\\ 
\textbf{Dice Loss:}
For a labeled set $\mathcal{D_L}$, the segmentation network is trained in a traditional supervised manner comprising dice loss, $L_{(seg)}^{\mathcal{L}} (x,y)= \sum_{x_i,y_i \in \mathcal{D_L} } {\mathcal{L}}_{dice}(x_i,y_i) = \sum_{x_i,y_i \in \mathcal{D_L} } \Big[1 - \frac{2 \sum_{x_j \in x_i, y_j \in y_i} f_1 (x_i) y_i}{\sum_{x_j \in x_i, y_j \in y_i} f_1 (x_j) + \sum_{y_j \in y_i} y_j} \Big]$. Then we define the supervised loss for distance map generation task as the mean squared error (MSE) loss between the predicted probability map $f_2(x)$ and the transformed ground truth map $\pi (y)$: $L_{(dis)}^{\mathcal{L}}(x,y)=\sum_{x_i,y_i \in \mathcal{D_L} } ||f_2(x_i) - \pi (y_i) ||$. \\
\textbf{Smoothing Loss:} We utilize a smoothing loss function $L_{(cross-task)}$ to enforce smoothness between the predicted segmentation mask and the inverse transform of the distance map as in \cite{luo2021semi}: $L_{(cross-task)}(x)=\sum_{x_i \in \mathcal{D} } ||f_1(x_i) - \pi^{-1} (f_2(x_i)) ||^2=\sum_{x_i \in \mathcal{D} } ||f_1(x_i) - \frac{1}{1+e^{-k.(f_2(x_i))}} ||^2$.\\
\textbf{Guidance Loss:}
As the uncertainty maps give the model some amount of interpretability with which we can decide whether the final segmentation is to be trusted, we consider using Monte-Carlo dropout (MC-dropout) \cite{kendall2017uncertainties} thanks to straightforward implementation. Voxel-wise segmentation uncertainty from MC dropout models is estimated as the mean entropy over all N samples generated by running inference on an input volume N times providing outputs with a set of probability vector of softmax scores, $\{P_n\}_{n=1}^N$ which captures a combination of aleatoric and epistemic uncertainty as, $U(x) = - \frac{1}{N} \sum_{i=1}^N p(x)log(p_i(x))$. We exploit the uncertainty as the guidance to filter out the high uncertainty (unreliable) predictions to minimize the voxel-level mean squared error (MSE) loss between the predicted mask and the transformed mask generated from the distance map: $\mathcal{L}_{G} =$ $\frac{\sum_{x_i \in (h\times w \times d)} \hat {\mathcal{B}} (U(x)<t) ||f_1(x_i) - \pi^{-1} (f_2(x_i)) ||^2}{\sum_{x_i \in  (h\times w \times d)} \hat {\mathcal{B}} (U(x)<t) }$; where $\hat {\mathcal{B(.)}}$ represents the indicator function for the uncertainty $U(x)$ with threshold $t$; $f_1(x)$ and $\pi^{-1} (f_2(x_i))$ are the prediction of main decoder and the distance map auxiliary decoder respectively.\\
\textbf{Adversarial-Geman-McClure Loss:}
On the other hand, the data with no corresponding segmentations are trained by minimising the unsupervised loss via a KL divergence which is based on LeastSquares-GAN. However, least-square loss is not robust. Instead, we adopt a new divergence loss function by incorporating it into a Geman-McClure model fashion called \textit{adversarial-Geman-McClure (adv-GM)} loss between the labeled data $x_{l}$ and the unlabeled data $x_{ul}$: 
\vspace{-2mm}
 %\cite{barron2019general}, 
\begin{equation} \label{eq:4}
\begin{split}
& L_{(adv-GM)}^{\mathcal{U}} ~=\\
&~ \frac{D\{x^{l}, dist_l ; \phi \}^2 + \{D(x^{ul}, dist_{ul} ; \phi)  - 1\}^2}{2\beta + D\{x^{l}, dist_l ; \phi \}^2 + \{D(x^{ul}, dist_{ul} ; \phi)  - 1\}^2};
\end{split}
\end{equation}

\noindent
where $dist_{ul} = f_{dis}(x^{ul}; \theta )$, $\beta$ is the scale factor which varies in the range of $[0,1]$ and we set $\beta=0.5$ in our experiment.

\noindent
\textbf{Data:} The model was trained and tested on the MICCAI STACOM 2018 Atrial Segmentation Challenge datasets featuring 100 3D gadolinium-enhanced MR imaging scans (GE-MRIs) and LA segmentation masks, with an isotropic resolution of $0.625 \times 0.625 \times 0.625mm^3$. The dimensions of the MR images may vary depending on each patient, however, all MR images contain exactly 88 slices in the $z$ axis. All the images were normalized and resized to $112 \times 112 \times 80$ before feeding them to the models. We split them into 80 scans for training and 20 scans for validation, and apply the same pre-processing methods.
\begin{figure*}[ht]
\centering
\includegraphics[width=0.99\linewidth]{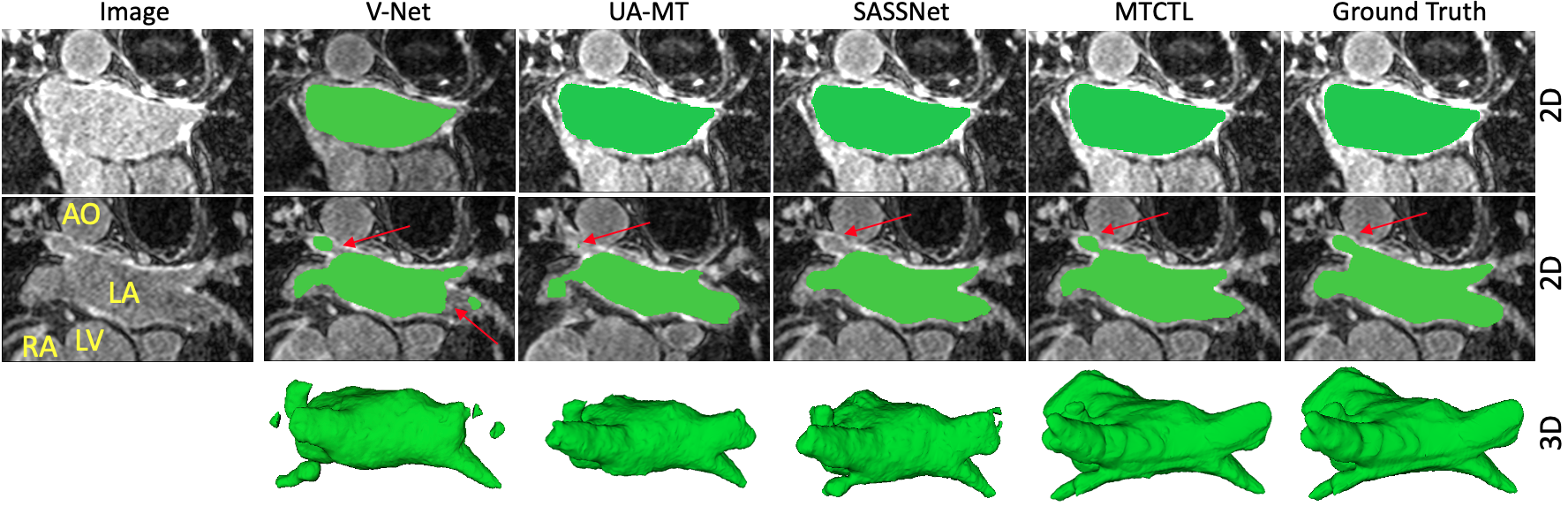}
\caption{Qualitative comparison of left atrium segmentation result in 2D as well as 3D of the MICCAI STACOM 2018 Atrial Segmentation challenge dataset yielded by four different frameworks: V-Net, UA-MT, SASSNet, and MTCTL. The comparison of segmentation results between the proposed method and three typical deep learning networks indicates that the performance of our proposed network is superior. Red arrow indicates the networks fail to capture the masks near Aorta (AO) region in 3D.}
\label{fig2}
\end{figure*}

\vspace{-3mm}
 \section{ Results and Discussion}
Figure \ref{fig2} shows the results obtained by V-Net \cite{milletari2016v}, UA-MT \cite{yu2019uncertainty}, SASSNet \cite{li2020shape}, our MTCTL, and the corresponding ground truth on the MICCAI STACOM 2018 Atrial Segmentation Challenge from left to right. The second row of the figure shows that all the three frameworks shows a portion of missing masks (red arrow) near Aorta (AO) region, whereas MTCTL generates more complete left atrium segmentation  following  the  addition of multiple tasks (distance map, cross-tasks, and uncertainty guidance) as multiple decoders in either 3D or 2D view. 

We conducted a paired statistical test to compare the segmentation performance in Table \ref{table1} which shows that our proposed model significantly improved the segmentation performance compared to the semi-supervised, fully-supervised, singletask, and multitask models in terms of the Dice, Jaccard, 95\% Hausdorff Distance (95HD), average surface distance (ASD), relative absolute volume difference (RAVD), Precision, and Recall. By exploiting unlabeled data with multiple tasks effectively, our proposed MTCTL model yielded a statistically  significant 7.2\%  improvement $(p<0.05)$ in Dice and  12.5\% Jaccard $(p<0.05)$ over the single tasked V-Net framework;  a  statistically significant 2.5\%improvement $(p<0.05)$ in Dice and 4.4\% Jaccard $(p<0.05)$ over the SASSNet framework with only 20\% labeled training data. 

\begin{figure}[H]
%\centering
\includegraphics[width=1.0\linewidth]{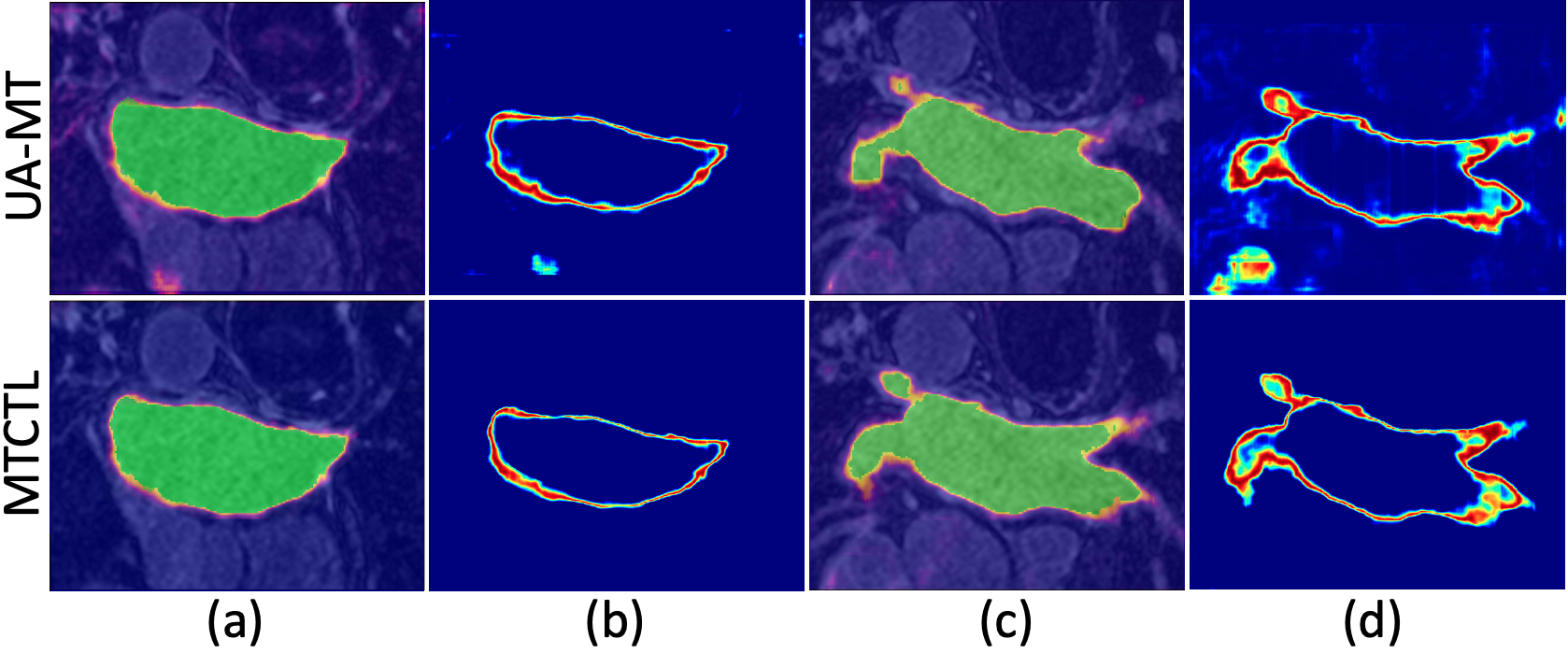}
\caption{ Visual comparison of segmentation predictions overlayed with uncertainty and uncertainty-only (predictive entropy) slices. Segmentation accuracy decreased while predictive uncertainty increased (low uncertainty shown in purple and high uncertainty shown in yellow). Segmentation mask overlaid with uncertainty ((a) $\&$ (c)), along with uncertainty maps ((b) $\&$ (d)) for two different slices of a patient. }
\label{fig3}
\vspace{-1mm}
\end{figure}

Figure \ref{fig3} shows a visual comparison of the uncertainty for segmentation of left atrium images in the coronal view. The first and second row presents the uncertainty over segmentation and the uncertainty only for two different slices obtained by the UA-MT and MTCTL framework respectively. In the uncertainty maps, blue pixels have low uncertainty values and red-ish pixels have high uncertainty values. It can be observed from the uncertainty map that the highest uncertainties are located near the border of the segmented foreground, while the pixels with a larger distance to the border have a very low uncertainty.

\begin{table*}[ht]
\scriptsize
\caption{Quantitative comparison of left atrium segmentation across several frameworks. Mean (standard deviation) values are reported for $Dice(\%), Jaccard(\%), 95HD(\%), ASD(\%), RAVD(\%), Precision(\%)$, and $Recall(\%)$ from all networks against our proposed MTCTL. The statistical significance of the results for MTCTL model compared against the baseline model SASSNET for $10\%$ and $20\%$ labeled data are represented by $*$ and $**$ for $p-$values $0.1$ and $0.05$, respectively. The best performance metric is indicated in \textbf{bold} text.}
\begin{center}
\begin{tabular}{p{2.2cm}>{\centering\arraybackslash}m{0.7cm}>{\centering\arraybackslash}m{0.9cm}||>{\centering\arraybackslash}m{1.3cm}>{\centering\arraybackslash}m{1.3cm}>{\centering\arraybackslash}m{1.4cm}>{\centering\arraybackslash}m{1.2cm}>{\centering\arraybackslash}m{1.2cm}>{\centering\arraybackslash}m{1.5cm}>{\centering\arraybackslash}m{1.1cm}}

\hline
\hline
\multirow{2}{1.4cm} & \multicolumn{2}{c|}{ \textbf{SCANS USED}} & \multicolumn{7}{c}{\textbf{METRICS}}\\

\cline{2-10}

\textbf{METHODS} & Labeled & Unlabeled & Dice(\%) $\uparrow$ & Jaccard(\%)$\uparrow$ & HD95(mm) $\downarrow$ & ASD(mm)$\downarrow$ & RAVD(\%) & Precision(\%) $\uparrow$ & Recall(\%)$\uparrow$ \\
\hline
V-Net \cite{milletari2016v} & 10\% & 0 & 79.98 $\pm$1.88  & 68.14$\pm$2.01  &  21.12$\pm$15.19 & 5.47$\pm$1.92 & -1.34$\pm$2.78  &  83.67$\pm$1.79 & 74.55$\pm$1.90 \\

UA-MT \cite{yu2019uncertainty}  & 10\% & 90\%& 84.25$\pm$1.61 & 73.48$\pm$1.73 & 13.84$\pm$13.15 & 3.36$\pm$1.58 & -0.13$\pm$2.56  & 87.57$\pm$1.53 & 77.85$\pm$1.65\\

SASSNet \cite{li2020shape} & 10\% & 90\% & 87.32$\pm$1.39 & 77.72$\pm$1.49 & 12.56$\pm$11.30 & 2.55$\pm$1.86 & -0.09$\pm$2.26  & 87.66$\pm$1.38 & 87.22$\pm$1.37\\
\textbf{MTCTL (Proposed)}  & 10\% & 90\% &  \textbf{*89.28$\pm$0.76} &  \textbf{*80.92$\pm$.79} &  \textbf{*7.74$\pm$6.05} &  \textbf{2.0$\pm$1.02} &  0.56$\pm$1.58  &  \textbf{*89.74$\pm$0.71} &  \textbf{*89.40$\pm$0.68 }\\

\hline

V-Net \cite{milletari2016v} & 20\% & 0 & 85.64$\pm$1.73  & 75.40$\pm$1.84  &  16.96$\pm$14.37 & 4.03$\pm$1.53 & -0.05$\pm$2.64 &  88.78$\pm$1.70 & 83.79$\pm$1.51 \\

UA-MT \cite{yu2019uncertainty} & 20\% & 80\% &  88.88$\pm$0.73 & 80.20$\pm$0.82 & 8.13$\pm$6.78 & 2.35$\pm$1.16  & -2.74$\pm$1.58  & 89.57$\pm$0.73 & 88.82$\pm$0.72\\

SASSNet \cite{li2020shape} & 20\% & 80\% & 89.54$\pm$\textbf{0.66}  &  81.24$\pm$\textbf{0.75} & 8.24$\pm$6.58 & 2.27$\pm$0.81 & 0.03$\pm$\textbf{1.55} & 89.86$\pm$\textbf{0.65} & 90.42$\pm$\textbf{0.66}\\

\textbf{MTCTL (Proposed)}  & 20\% & 80\% &  \textbf{**91.80}$\pm$0.67 & \textbf{**84.80}$\pm$0.83 & \textbf{**5.50}$\pm$4.74 & \textbf{1.55}$\pm$0.28  & \textbf{0.01}$\pm$1.65  & \textbf{91.15}$\pm$0.76 & \textbf{91.04}$\pm$0.75\\
\hline 
\hline
\end{tabular}
\label{table1}
\end{center}
\end{table*}

\section{Conclusion}
In  this  paper,  we  proposed  a  multi-task cross-task learning network (MTCTL) for atrial segmentation. To improve robustness beyond that of  the  recent SOTA  framework, we utilize model uncertainty derived from Monte Carlo Sampling to serve as local guidance between the predicted segmentation mask and the mask generated by transforming the distance map. Our enforced cross-task loss correlates between the pixel-level (segmentation) and the geometric-level (distance map) tasks to generate smoother and more accurate segmentation masks. We evaluated its performance on the MICCAI STACOM 2018 Atrial Segmentation Challenge dataset.  We also  conducted  an “uncertainty” estimation analysis to determine where our algorithm “fails” to segment regions of interest in an image. Our   proposed   model outperforms  existing  methods  in  terms  of  both  Jaccard and Dice, achieving 89.3\% Dice and 80.9\% Jaccard with only 10\% labeled data and 91.8\% Dice and 84.8\% Jaccard with only 20\% labeled data for atrial segmentation, both of which showed at least 2.5\% improvement over the best methods and  more  than  7\%  improvement over single-task traditional V-Net architecture. 

To our best knowledge, the proposed MTCTL framework constitutes the first approach to adopt adversarial approach along with uncertainty estimation and most accurate semi-supervised left atrium segmentation performance on the LA database. As part of future work, we will use 
these uncertainty maps to detect regions where the segmentation of the left and right ventricle myocardium and blood pool fails, which is a critical feature for both research and clinical applications.

\section*{Acknowledgments}  
This work was supported by grants from the National Science Foundation (Award No. OAC 1808530) and the National Institutes of Health (Award No. R35GM128877).

\bibliography{cinc}

\end{document}